\documentclass[a4paper,11pt]{article}
\usepackage{jheppub} 
\usepackage[T1]{fontenc} 
\usepackage{amssymb}
\usepackage{amsmath}
\usepackage{float} 
\usepackage{subfigure}
\usepackage{booktabs}
\usepackage{mathrsfs}

\title{Non-Topological Soliton Bardeen Boson Stars and Their Frozen States}

\author[*,a]{  Zhen-Hua Zhao}
\author[a]  {, Yi-Ning Gu }
\author[b]  {, Shu-Cong Liu }
\author[b]  {, Long-Xing Huang }
\author[*,b]{, Yong-Qiang Wang }
\affiliation[a]{Department of Applied Physics, Shandong University of Science and Technology, Qingdao 266590,  China\\
}
\affiliation[b]{
School of Physical Science and Technology, Lanzhou University, Lanzhou 730000, China}

\emailAdd{zhaozhh78@sdust.edu.cn, yqwang@lzu.edu.cn, $^*$corresponding author}

\abstract{

We investigate a Bardeen model coupling Einstein gravity with nonlinear electromagnetic fields and non-topological soliton complex scalar fields, governed by the magnetic charge $\tilde{q}$, the complex scalar field frequency $\tilde{\omega}$, and the self-interaction parameter $\tilde{\eta}$. Our results reveal that the magnetic charge $\tilde{q}$ exhibits $\tilde{\eta}$-dependent critical values $\tilde{q}_c$, beyond which ($\tilde{q} > \tilde{q}_c$) Bardeen boson stars (BBSs) may transition into frozen states ($\tilde{\omega} \to 0$). These frozen states are characterized by a critical horizon whose radius $\tilde r^\mathrm{H}_{c}$ satisfies $\tilde r_{\text{inner}}^{\text{H,RN}} < \tilde r_{c} < \tilde r_{\text{outer}}^{\text{H,RN}}$, where $\tilde r_{\text{inner}}^{\text{H,RN}}$ and $\tilde r_{\text{outer}}^{\text{RN}}$ denote the inner and outer horizons of magnetic Reissner-Nordstr\"{o}m (RN) black holes with equivalent mass and magnetic charge. Notably, the ADM mass  of frozen BBSs is independent of $\tilde{\eta}$. Furthermore, light ring (LR) solutions exist universally across all tested  combinations of $\tilde{q}$ and $\tilde{\eta}$, with all frozen BBSs exhibiting LRs whose outer radius $\tilde r_{\text{outer}}^{\text{LR}}$ is independent of $\tilde{\eta}$. Compared to magnetic RN black holes, frozen BBSs possess a smaller outer LR radius ($\tilde r_{\text{outer}}^{\text{LR}} < \tilde r_{\text{outer}}^{\text{LR, RN}}$).

}

\begin{document} 
\maketitle
\flushbottom

\section{Introduction}\label{sec:intro}

The existence of black holes has been reinforced by the discovery of gravitational waves produced by the merger of two black holes \cite{AbbottAbbottAbbott2016} and the observation of black hole shadows by the Event Horizon Telescope (EHT) \cite{AkiyamaAlberdiAlef2019e,AkiyamaAlberdiAlef2019a,AkiyamaAlberdiAlef2022e}. The realisation of these experiments represents a significant milestone in the process of black hole research. However, there remain a number of long-standing issues that require further investigation, including the existence of singularities and singular rings, as well as the so-called Black hole information paradox \cite{Hawking1974,Wald1975,Hawking1976}. Penrose's singularity theorem \cite{Penrose1965} led to the understanding that the existence of a singularity in the solution of a black hole in general relativity appeared to be an intrinsic feature of the theory.  Although the weak cosmic censorship conjecture \cite{Penrose1969} permits singularities of gravitational collapse to be contained within the event horizon, which effectively shields them from external observers, this does not alter the fundamental reality of the existence of singularities.

The existence of a singularity implies that matter can be compressed to a point, which may be acceptable as a mathematical object, but challenging to accept as a physical entity \cite{Einstein1939,Kerr2023}.
The study of the nature of the singularity has also prompted interest in the study of  singularity-free black holes, known as regular black holes \cite{Bardeen1968a,BronnikovMelnikovShikin1979,Borde1994,BarrabesFrolov1996,MarsMartin-PratsSenovilla1996,Borde1997,CaboAyon1999,Hayward2006,FanWang2016,Fan2017}. The earliest known study in this field was proposed by Bardeen in 1968 \cite{Bardeen1968a}, who suggested a metric for a black hole without singularities, which is now known as the ``Bardeen black holes'' \cite{Borde1997}. However, obtaining exact solutions for singularity-free black holes by solving the Einstein equations remains highly challenging.  It was not until 1999 that Ayón-Beato and García \cite{Ayon-BeatoGarcia1998} proposed the first exact regular black hole solution using nonlinear electrodynamics coupled to general relativity. Subsequently, they also derived the exact solution for the Bardeen black hole \cite{Ayon-BeatoGarcia2000}, interpreting it as a nonlinear magnetic monopole.

The construction of general relativistic objects by introducing electromagnetic fields and Einstein gravitational coupling can be traced back to the work of Wheeler in the 1950s, who termed this ``gravitational-electromagnetic entity'' as {\it geons} \cite{Wheeler1955}. In the 1960s, Kaup \cite{Kaup1968} suggested massive complex scalar fields as an alternative to the electromagnetic field. The stable local solutions that resulted from this were known as ``Klein-Gordon Geon'' and subsequently as boson stars (BSs).

Unlike black holes, BSs lack event horizons. However, recent studies  \cite{Wang2023,YueWang2023a} have shown that, under specific conditions, BSs can possess an critical horizon   that closely resembles the event horizon of black holes. From the perspective of an observer at infinity, time within the critical horizon appears to slow down infinitely, causing matter to asymptotically approach but never cross the horizon within a finite time. BSs in this state are called ``frozen stars" . The concept of frozen stars resulting from gravitational collapse in a vacuum was first introduced by Oppenheimer and Snyder \cite{OppenheimerSnyder1939}.

 The frozen stars studied in Ref.~\cite{Wang2023} are are constructed within the framework of the Bardeen model, where Einstein gravity is coupled to a nonlinear electromagnetic field and a free complex scalar field. In this paper, based on Wang's work~\cite{Wang2023}, we  further explore   scenarios involving  self-interactions of the complex scalar field.    Specifically, we focus on the case where the complex scalar field adopts a non-topological configuration. Its coupling with Einstein gravity can give rise to a non-topological soliton star, a concept first proposed by Friedberg, Lee, and Pang \cite{Lee1987,FriedbergLeePang1987a}) and subsequently studied extensively in, for example, Refs.~\cite{Lynn1989,Jetzer1992a,KleihausKunzList2005,MacedoPaniCardoso2013,DzhunushalievFolomeevHoffmann2014,BrihayeCisternaHartmann2015, Macedo2015,CollodelKleihausKunz2017,BoskovicBarausse2022,CollodelDoneva2022, Siemonsen2024,OgawaIshihara2024}.


The organization of this paper is as follows. In Sec. \ref{model}, we present the  model what we discussed. In Sec.  \ref{sect3}, we outline Numerical  Scheme. In Sec.  \ref{sect4}, we present the numerical results and the discussion  in four topics. In Sec. \ref{Conc}, we present some conclusions.

\section{The model}\label{model}
This section provides a brief introduction to the theoretical framework, which includes the Einstein nonlinear electrodynamics model \cite{Ayon-BeatoGarcia2000} coupled to a self-interaction complex scalar field.  We adopt the natural units where $\hbar = c =1$. The action is given by follows:
\begin{equation}\label{action}
  S=\int\sqrt{-g}d^4x\left(\frac{R}{16\pi { G}}+\mathcal{L}^{(1)}+\mathcal{L}^{(2)}\right), 
\end{equation}
where the Lagrangian densities are defined as
\begin{eqnarray}
\mathcal{L}^{(1)} &= &-\frac{3}{2s}\Bigg(\frac{\sqrt{2q^2\mathcal{F}}}{1+\sqrt{2q^2\mathcal{F}}}\Bigg)^{\frac{5}{2}},\\
\mathcal{L}^{(2)} &= & -\nabla_\mu \Phi^* \nabla^ \mu \Phi  - U(\Phi,\Phi^*).
\end{eqnarray}
Here,
\begin{equation}\label{Pot1}
U(\Phi\Phi^*) = \mu^2 \Phi \Phi^* \left(1-2 \eta^2\Phi\Phi^*\right)^2,
\end{equation}
 $R$ is the scalar curvature, $\Phi$ is a complex scalar field, $\Phi^*$ is the complex conjugate of $\Phi$,  ${\cal F} = \frac{1}{4}F_{\mu\nu} F^{\mu\nu}$ with the electromagnetic field strength $F_{\mu\nu} = \partial_{\mu} A_{ \nu} - \partial_{\mu} A_{ \nu}$, and $A_{\mu}$ represents the electromagnetic field. The constants $q$, $s$, $\mu$, and $\eta$ are independent parameters, where $q$ denotes the magnetic charge, $\mu$ the scalar field mass parameter, and $\eta$ the coupling constant of the scalar fields.  
 
When $q = 0$ while the complex scalar field remains non-vanishing, the action describes a non-topological soliton black hole (BS) theory. Conversely, if the complex scalar field vanishes but $q \neq 0$, the model reduces to the Bardeen black hole theory. The line element of the Bardeen theory in spherical coordinates $(t, r, \theta, \varphi)$ is given by

\begin{equation}
ds^2=-f(r)dt^2+f(r)^{-1}dr^2+r^2\left(d\theta^2+\sin^2\theta d\varphi^2\right),
\end{equation}
where
\begin{equation}\label{eqfr} 
f(r)=1-\frac{q^3r^2 { 4\pi G}}{s(r^2+q^2)^{3/2}}.
\end{equation}
For $q \geq q_b = 3^{3/4} \sqrt{s/({ 8\pi G}})$, the equation $f(r) = 0$ has a real root, indicating a Bardeen spacetime solution with an event horizon. When $q < q_b$, there is no event horizon, and thus no black hole solution. The asymptotic behavior of $f(r)$ at infinity is
\begin{equation}
f(r)=1-2\frac{q^3 {\color{blue}4\pi G}}{ 2 s }\frac{1}{r}+\mathcal{O}(1/r^3).
\end{equation}
The Arnowitt-Deser-Misner (ADM) mass $M$ of this model, derived from the second term, is given by $ { M = q^3 4\pi / (2s) }$ \cite{Carroll2014a}.

Varying the action \eqref{action} with respect to $g_{\mu\nu}$, $A_{\mu}$, and $\Phi$ yields the equations of motion:
\begin{eqnarray} \label{eq:EKG1}
R_{\mu\nu}-\frac{1}{2}g_{\mu\nu}R- 8 \pi { G} (T^{(1)}_{\mu\nu}+T^{(2)}_{\mu\nu})&=&0 \ ,  \\
\nabla_{\mu} \left(\frac{ \partial {\cal L}^{(1)}}{ \partial {\cal F}}  F^{\mu\nu}\right) &=& 0,    \\
\Box\Phi- \dot{U}(\Phi\Phi^*)\Phi &=& 0. 
\end{eqnarray}
Here,
\begin{equation}
\dot{U}(\Phi\Phi^*) = \frac{\partial U(\Phi,\Phi^*)}{\partial  (\Phi\Phi^*)}   =  \mu^2 \left[1-4 \eta^2 (\Phi\Phi^*)+3 \eta^4 (\Phi\Phi^*)^2\right] ,
\end{equation}
\begin{equation}
T^{(1)}_{\mu\nu} =- \frac{ \partial {\cal L}^{(1)}}{ \partial {\cal F}} F_{\mu \rho} F_{ \nu }^{\;\;\rho} + g_{\mu\nu} {\cal L}^{(1)},
\end{equation}
\begin{equation}
T^{(2)}_{\mu\nu} = \partial_\mu \Phi^* \partial_\nu \Phi + \partial_\nu \Phi^* \partial_\mu \Phi - g_{\mu\nu} \left[ \frac{1}{2}g^{\lambda\rho}\left(\partial_\lambda\Phi^*\partial_\rho\Phi + \partial_\rho\Phi^*\partial_\lambda\Phi\right) + U(\Phi,\Phi^*)\right]\,.
\end{equation}

The action \eqref{action} is invariant under the global $U(1)$ transformation of the complex scalar field, $\Phi \rightarrow e^{i\alpha} \Phi$, where $\alpha$ is a constant. According to Noether's theorem, this symmetry leads to a conserved charge:

\begin{equation}\label{equ9}
	Q = \int_{\varSigma}J^0 \mathrm{d} V,  
\end{equation}
where $\varSigma$ is a spacelike hypersurface and $J^0$ is the timelike component of the conserved current:
\begin{equation}\label{equ8}
	J^{\mu} = -i\left(\Phi^*\partial^\mu\Phi - \Phi\partial^\mu\Phi^*\right).
\end{equation}

We consider the generic static, spherically symmetric spacetime with the line element
\begin{equation}\label{equ10}
	ds^2 = -N(r)\sigma^2(r)dt^2 + \frac{dr^2}{N(r)} + r^2\left(d\theta^2 + \sin^2\theta d\varphi^2\right),
\end{equation}
where $N(r) = 1 - 2  { G} m(r)/{r}$. The functions $N(r)$, $m(r)$, and $\sigma(r)$ depend only on the radial coordinate $r$. The ansatzes for the electromagnetic field \cite{Ayon-BeatoGarcia2000} and the scalar field \cite{Lee1987,FriedbergLeePang1987a} are given by
\begin{equation}\label{equ11}
  F_{\mu\nu}=2\delta_{[\mu}^\theta\delta_{\nu]}^\varphi q \sin(\theta),\;\;\; \Phi = \frac{\phi(r)}{\sqrt{2}} e^{-i\omega t},
\end{equation}
where $\phi(r)$ is a real scalar field and $\omega$ is the frequency of $\Phi$.

The Noether charge obtained from  Eq.~(\ref{equ9}) is written as
\begin{equation}\label{equ18}
	Q = 8 \pi   \int_0^\infty r^2\frac{\omega\phi^2}{N~\sigma}dr\, .
\end{equation}

Substituting the ansatzes (\ref{equ10}) and (\ref{equ11}) into the field equations (\ref{eq:EKG1}), we obtain the following equations:

\begin{eqnarray}
	 \phi^{\prime\prime}+\left(\frac{2}{r} + \frac{N^\prime}{N} + \frac{\sigma^\prime}{\sigma}\right)\phi^\prime + \left(\frac{\omega^2}{N \sigma^2} - \dot{U}(\phi)\right)\frac{\phi}{N} &=& 0\, ,\label{ode11}\\
 N' +4 \pi { G} \frac{\omega^2 r \phi^2 }{N \sigma^2}+  4 \pi  { G}  N r  \phi'^2+\frac{N}{r} + r 8 \pi { G} U(\phi) 
+  \frac{12 \pi   { G} r}{s \left(1+r^2/q^2\right)^{5/2}}-\frac{1}{r}&=&0, \label{ode12}
 \\
  	\frac{\sigma^\prime}{\sigma} - 4 \pi { G}  r\left(\phi^{\prime2} + \frac{\omega^2\phi^2}{N^2 \sigma^2}\right)&=&0,  \label{ode13}
\end{eqnarray}
where
$$
\begin{aligned}
{U}(\phi) & =  \frac{\mu^2}{2} \phi^2\left(1- \eta^2 \phi^2 \right)^2, \\
\dot{U}(\phi) &=\frac{\partial {U}(\phi) }{\partial (\phi^2)} =  \mu^2 \left(1-4 \eta^2 \phi^2+3 \eta^4 \phi^4 \right), \\
 N' &= -\frac{2  { G} m'(r)}{r}+\frac{2  { G} m(r)}{r^2},
\end{aligned}
$$
and the prime denotes differentiation with respect to $r$. {  Here, we assume that all functions appearing in Equations \eqref{ode11}-\eqref{ode13} are differentiable over the range from zero to infinity}.

To solve these ordinary differential equations, we impose the following boundary conditions to ensure regularity at the origin and asymptotic flatness at infinity:
\begin{equation}\label{equ19}
m(0) = 0, \  \ \sigma(\infty) = 1,  \ \  m(\infty) =  M , \ \  \sigma(0) = \sigma_0.
\end{equation}
where $M$ is the ADM mass, and $\sigma_0$ is a constant determined by solving the differential equations. The scalar field boundary conditions are:

\begin{equation}\label{equ20}
\phi(\infty) = 0,\;\;\;\left. \frac{d\phi(r)}{dr}\right|_{r = 0} = 0.
\end{equation}

\section{Numerical  scheme} \label{sect3}

To facilitate numerical computations, we apply the following scaling transformations:

\begin{eqnarray}\label{scaling}
\   r\to \frac{\tilde{r}}{ \mu}, \ q \to  \frac{\tilde{q}}{ \mu}, \ \eta \to  {\tilde{\eta} }{ \sqrt{4\pi  { G}}}, \ \omega \to \tilde{\omega}{\mu},\ s \to  \frac{4 \pi { G} \tilde{s}}{ \mu^2}, \ m \to \frac{\tilde{m} }{  { G} \mu}, \  \phi \to \frac{\tilde{\phi} }{ \sqrt{4\pi  { G}}},\\
 \   M \to \frac{\tilde{M}}{ { G} \mu}, \   \mathrm{and}\  Q \to \frac{\tilde{Q}}{ { G} \mu^2},
\end{eqnarray}
where
\begin{equation}\label{equ18}
	\tilde{Q} =2 \int_0^\infty \tilde r^2\frac{\tilde\omega \tilde\phi^2}{N~\sigma} \mathrm{d} \tilde r\, .
\end{equation}

After applying these transformations,  { all physical quantities denoted by a tilde symbol are dimensionless, and } the equations \eqref{ode11}--\eqref{ode13} simplify by setting:

\begin{equation}
\mu \to 1, \ { G} \to 1 \ \mathrm{and}\   4\pi   \to 1.
\end{equation}
This demonstrates that the equations are scale-invariant with respect to the mass parameter $\mu$.

To further simplify the numerical implementation, we introduce a compactified radial coordinate:
\begin{equation}\label{x_r}
x=\frac{\tilde r}{1+\tilde r}.
\end{equation}
This transformation maps the domain $[0, \infty)$ to $[0, 1]$, with $x = 0$ corresponding to the origin and $x = 1$ to spatial infinity.

 The open-source software Fenicsx (version 0.8) \cite{BarattaDeanDokken2023,ScroggsEtal2022,BasixJoss,AlnaesEtal2014} is employed to solve the differential equations \eqref{ode11}, \eqref{ode12}, and \eqref{ode13} subject to the boundary conditions \eqref{equ19} and \eqref{equ20}. The computational domain $[0, 1]$ is discretized using at least 3000 mesh  points, ensuring a relative numerical error of less than $10^{-5}$.


 \section{Numerical results}\label{sect4}
 
\begin{figure}[htbp]

\subfigure[]{
\centering
\includegraphics[width=0.47\linewidth]{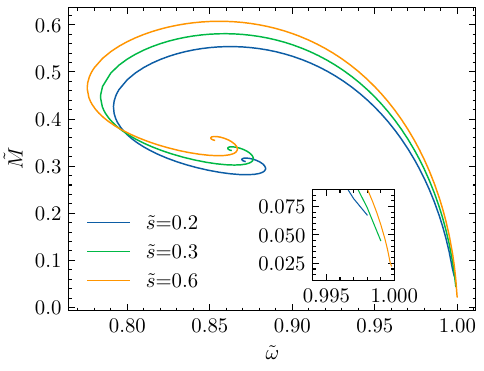} \label{M_s}
}
\subfigure[]{
\centering

\includegraphics[width=0.47\linewidth]{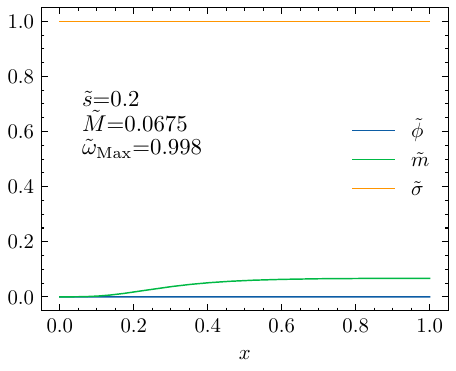} \label{omegamax}
}

\caption{(a) Effect of parameter $\tilde{s}$ on the $\tilde M$-$\tilde \omega$ relation; (b) Solutions behavior in the $\tilde{\omega}\to\tilde{\omega}_\mathrm{Max}$ limit. Parameters $\tilde{q}=0.2$ and $\tilde\eta=0$.}
\end{figure}

 In this section, we will present our numerical results. { We first discuss the influence of parameter $\tilde{s}$ on the solutions. As the frequency parameter $\tilde{\omega}$ approaches its maximum value $\tilde{\omega}_\mathrm{Max}$, the scalar field $\tilde{\phi}\to 0$ and $\tilde{\sigma}\to 1$, causing the metric function ${N}$ to asymptotically approach the Bardeen black hole solution. Under this limiting condition, the ADM mass admits an analytical expression
\begin{equation}\label{tildeM}
\tilde{M} = \frac{\tilde{q}^3}{2\tilde{s}}.
\end{equation}
If the magnetic charge $\tilde{q}$ is fixed, increasing $\tilde{s}$ reduces the  value of  $\tilde{M}$.

When the system deviates from the $\tilde{\omega} \to \tilde{\omega}_\mathrm{Max}$ limit, the parameter coupling manifests greater complexity. In such cases, the relationship among $\tilde{M}$, $\tilde{q}$, and $\tilde{s}$ no longer conforms to the simple form in Eq. \eqref{tildeM}, but is fundamentally governed by the coupled differential equations \eqref{ode11}-\eqref{ode13}.

Fig. \ref{M_s} displays numerical results for $\tilde{q}=0.3$, $\tilde\eta=0$, and $\tilde{s}\in\{0.2,0.3,0.6\}$. Remarkably, in the $\tilde{\omega}\to\tilde{\omega}_\mathrm{Max}$ regime, increasing $\tilde{s}$ indeed decreases $\tilde{M}$, in precise agreement with the analytical solution \eqref{tildeM}. However, when $\tilde{\omega}$ moves away from $\tilde{\omega}_\mathrm{Max}$, this parameter dependence reverses, with larger $\tilde{s}$ values  inducing larger $\tilde{M}$ values.

Fig. \ref{omegamax} further verifies theoretical predictions under extreme conditions: for $\tilde{q}=0.3$, $\tilde\eta=0$, and $\tilde{s}=0.2$, the numerical solution at $\tilde{\omega}\to\tilde{\omega}_\mathrm{Max}$ satisfies $\tilde{\phi}\to 0$ and $\tilde{\sigma}\to 1$ exactly. Crucially, the computed $\tilde{M}$ value shows complete consistency with the theoretical prediction from Eq. \eqref{tildeM} up to four decimal places.

Given that both $\tilde{q}$ and $\tilde{s}$ are dimensionless free parameters, we select $\tilde{s}=0.2$ as the typical value for subsequent analysis. This choice facilitates focused investigation into how the magnetic charge $\tilde{q}$, interaction parameter $\tilde{\eta}$, and frequency $\tilde{\omega}$ influence the system's characteristics.

}

 When $\tilde{\eta} = 0$, the model reduces to the one studied in Ref.~\cite{Wang2023}. It is known that when the value of $\tilde{q}$ exceeds a critical value, denoted as $\tilde{q}_c$, a solution with frequency $\tilde{\omega} \to 0$ exists. In our numerical calculations, the smallest value of $\tilde{\omega}$ is set to $0.0001$. However, in principle, solutions with even smaller values of $\tilde{\omega}$ can be obtained { by refining the mesh   (increasing the number of points)  and decreasing the step size of $\tilde \omega$.}

The model under consideration includes a self-interaction term for the complex scalar field. Our calculations reveal that the critical value $\tilde{q}_c$ varies with the self-coupling constant $\tilde{\eta}$. The values of $\tilde{q}_c$ for $\tilde{\eta} = 0.0, 0.5, 1.0, 1.5, 2.0, 2.5, 3.0, 3.5, 4.0$ are computed and summarized in Table~\ref{tab2}. From the table, we observe that the variation in $\tilde{\eta}$ has a subtle but non-negligible effect on $\tilde{q}_c$, with the maximum difference between $\tilde{q}_c$ values not exceeding $0.004$. Although this difference is small, its implications cannot be overlooked and will be further discussed in Subsection~\ref{sub412}.

 \begin{table}[tbp]
\centering
\begin{tabular}{cccccccccc}
\toprule
 $\tilde \eta $  &  0.0 &  0.5& 1.0& 1.5 & 2.0& 2.5&3.0 & 3.5& 4.0\\
\midrule
$\tilde q_c$  & 0.59979 & 0.60015& 0.60317 & 0.6028 & 0.6021& 0.60165 & 0.60136 &0.60115&0.60098\\

\bottomrule
\end{tabular}
\caption{The relation between  $\tilde \eta$ and $\tilde q_c$. There are different values of $\tilde q_c$ for different values of $\tilde \eta$.} \label{tab2} 
\end{table}

\subsection{Solutions}

\begin{figure}[htbp]

\subfigure[]{
\centering
\includegraphics[width=0.47\linewidth]{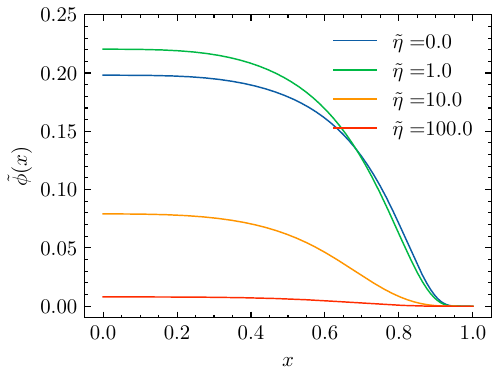} \label{ q=0.3phi_part}
}
\subfigure[]{
\centering

\includegraphics[width=0.47\linewidth]{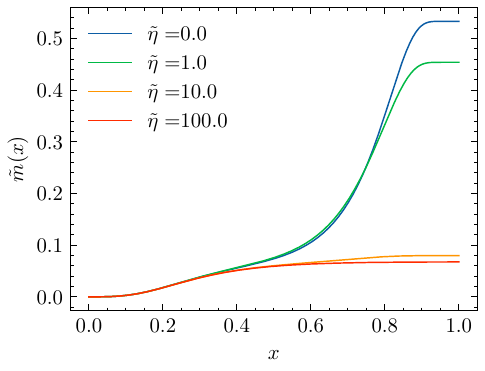} \label{ q=0.3M_part}
}

\centering
\subfigure[]{
\centering
\includegraphics[width=0.47\linewidth]{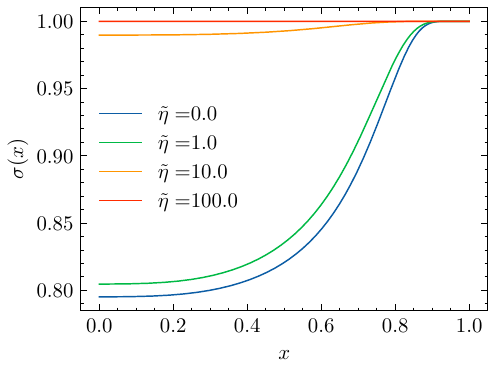} \label{ q=0.3sigma_part}
}
\caption{Solutions of  the fields functions with $\tilde q=0.3$, $\tilde \omega = 0.9$ and different values of $\tilde \eta$.
  }\label{sols_eta}
\end{figure}

In general, if the values of $\tilde q$ and $\tilde \omega$ are fixed, the solutions of the equations \eqref{ode11}, \eqref{ode12}, and \eqref{ode13} can cover a wide range of values of $\tilde \eta$. For example, $\tilde \eta$ can take values in the range $[0, 100]$. We present some representative solutions in Fig.~\ref{sols_eta}, with  $\tilde q=0.3$ and $\tilde \omega=0.9$. 

\subsubsection{Case of $\tilde q < \tilde q_c$}

\begin{figure}[htbp]
\centering
\subfigure[]{
\includegraphics[width=0.47\linewidth]{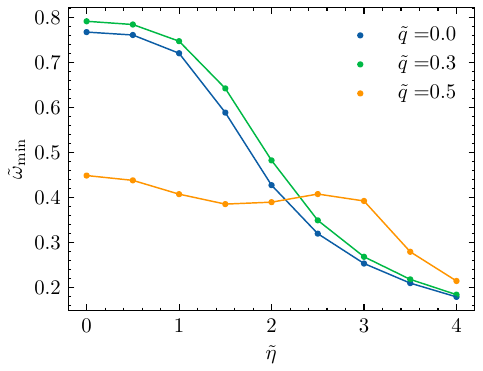} 
\label{2a}
}
\subfigure[]{
\includegraphics[width=0.47\linewidth]{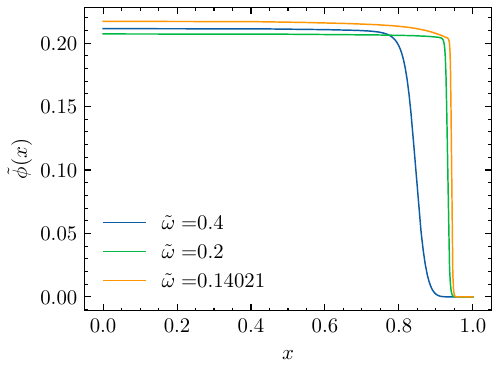} 
\label{2b}
}
\subfigure[]{
\includegraphics[width=0.47\linewidth]{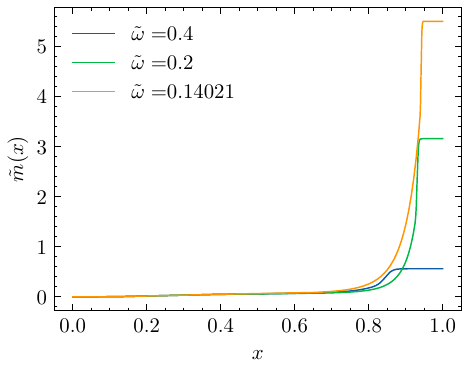} 
\label{2c}
}
\subfigure[]{
\includegraphics[width=0.47\linewidth]{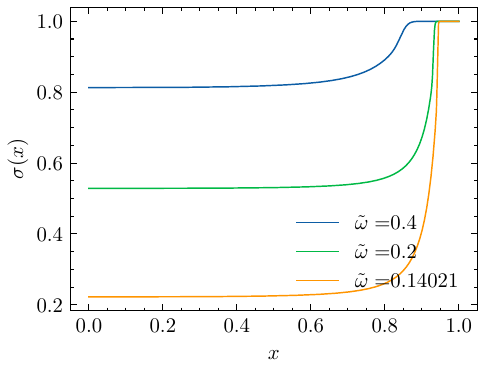} 
\label{2d}
}
\caption{(a) $\tilde \omega_\mathrm{min}$ versus $\tilde\eta$. (b)-(d) Solutions of the field functions with $\tilde\eta = 5$ and $\tilde q = 0.3$. }\label{eta=5}
\end{figure}

In Fig.~\ref{2a}, we illustrate the relationship between the minimum value of $\tilde \omega$, denoted by $\tilde \omega_\mathrm{min}$, and $\tilde\eta$. As $\tilde\eta$ increases, $\tilde \omega_\mathrm{min}$ generally decreases. Additionally, as $\tilde \omega$ decreases, the problem becomes increasingly stiff and more challenging to solve. To demonstrate this, the solutions for $\tilde\eta = 5$ are presented in Fig.~\ref{2b}, Fig.~\ref{2c}, and Fig.~\ref{2d}. Notably, the slope of the solutions for $\tilde\phi(x)$, $\tilde m(x)$, and $\sigma(x)$ around $x=0.95$ increases significantly as $\tilde \omega$ decreases.

\begin{figure}[htbp]
\centering
\subfigure[]{
\label{M-w-1}		
\includegraphics[width=0.47\linewidth]{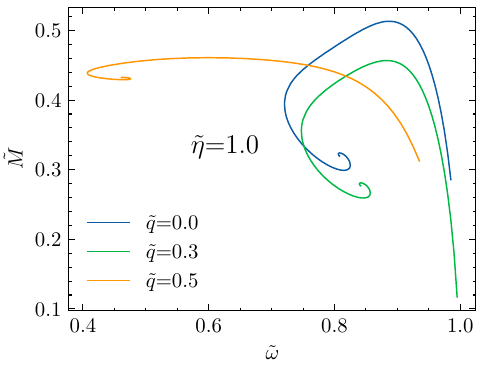}}
\subfigure[]{
\label{M-w-2}
\includegraphics[width=0.47\linewidth]{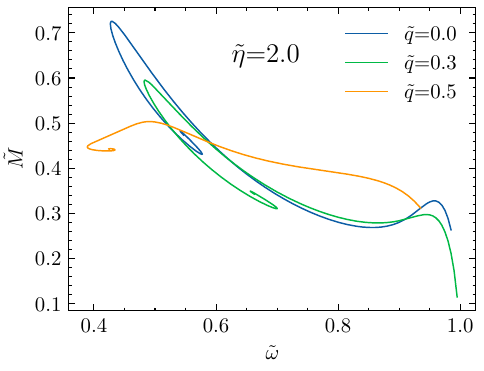}
}
\subfigure[]{
\label{M-w-3}
\includegraphics[width=0.47\linewidth]{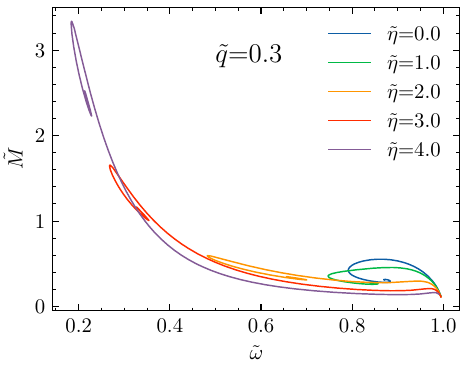}}
\subfigure[]{
\includegraphics[width=0.47\linewidth]{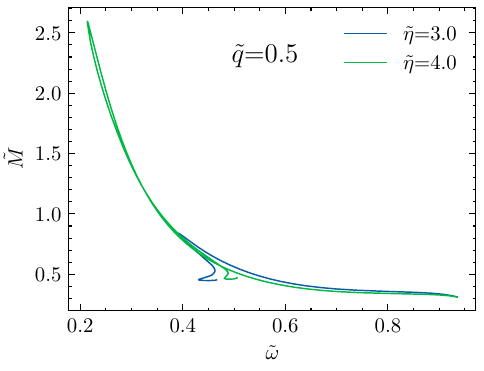}
\label{M-w-4}
}
\caption{ $\tilde M$ versus $\tilde \omega$ with $\tilde q < \tilde q_c$. }\label{Mvsw}
\end{figure}

When $\tilde \eta$ is relatively small, for instance, when $\tilde \eta=1$,  the $\tilde M$ versus $\tilde \omega$ curves maintain a classical spiral shape, as depicted in Fig.~\ref{M-w-1} with $\tilde q= 0.0,0.3,0.5$. As $\tilde \eta$ increases, such as when $\tilde \eta=2$, the curve exhibits a bimodal structure, as shown in Fig.~\ref{M-w-2} with $\tilde q =0.0,0.3$. Furthermore, there is a significant increase in the maximum value of $\tilde M$ as $\tilde \eta$ increases, as illustrated in Fig.~\ref{M-w-3}. As both $\tilde q$ and $\tilde\eta$ increase, the $\tilde M$ versus $\tilde \omega$ curves may adopt a very different shape from the classical one, for example, as shown in Fig.~\ref{M-w-4}, where cusps appear at the left end of the curves, and the curves almost fold into each other. Additionally, the lower branch of the curve intersects with the upper branch when $\tilde q=0.5$ and $\tilde \eta=4$.

\begin{figure}[htbp]
\centering
\subfigure[]{
\label{Q-w-1}
\includegraphics[width=0.47\linewidth]{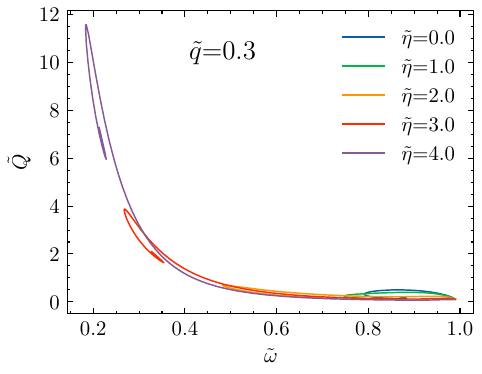}}
\subfigure[]{
\label{Q-w-2}
\includegraphics[width=0.47\linewidth]{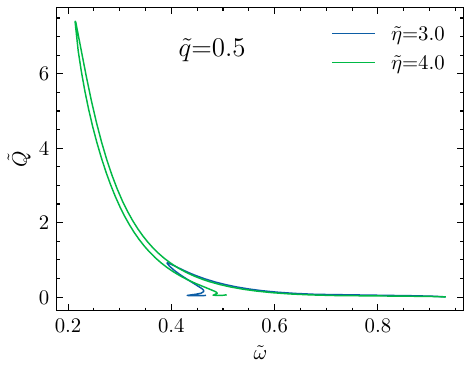}}
\caption{ $\tilde Q$ versus $\tilde \omega$ with $\tilde q < \tilde q_c$. }\label{Mvsw}
\end{figure}

The $\tilde Q$ versus $\tilde \omega$ curves exhibit similar characteristics to the $\tilde M$ versus $\tilde \omega$ curves, as shown in Fig.~\ref{Q-w-1}. Notably, the upper and lower branches of the curve do not cross when $\tilde q=0.5$ and $\tilde \eta=4$, as illustrated in Fig.~\ref{Q-w-2}.

\subsubsection{Case of $\tilde q > \tilde q_c$} \label{sub412}

\begin{figure}

\centering
\includegraphics[width=\linewidth]{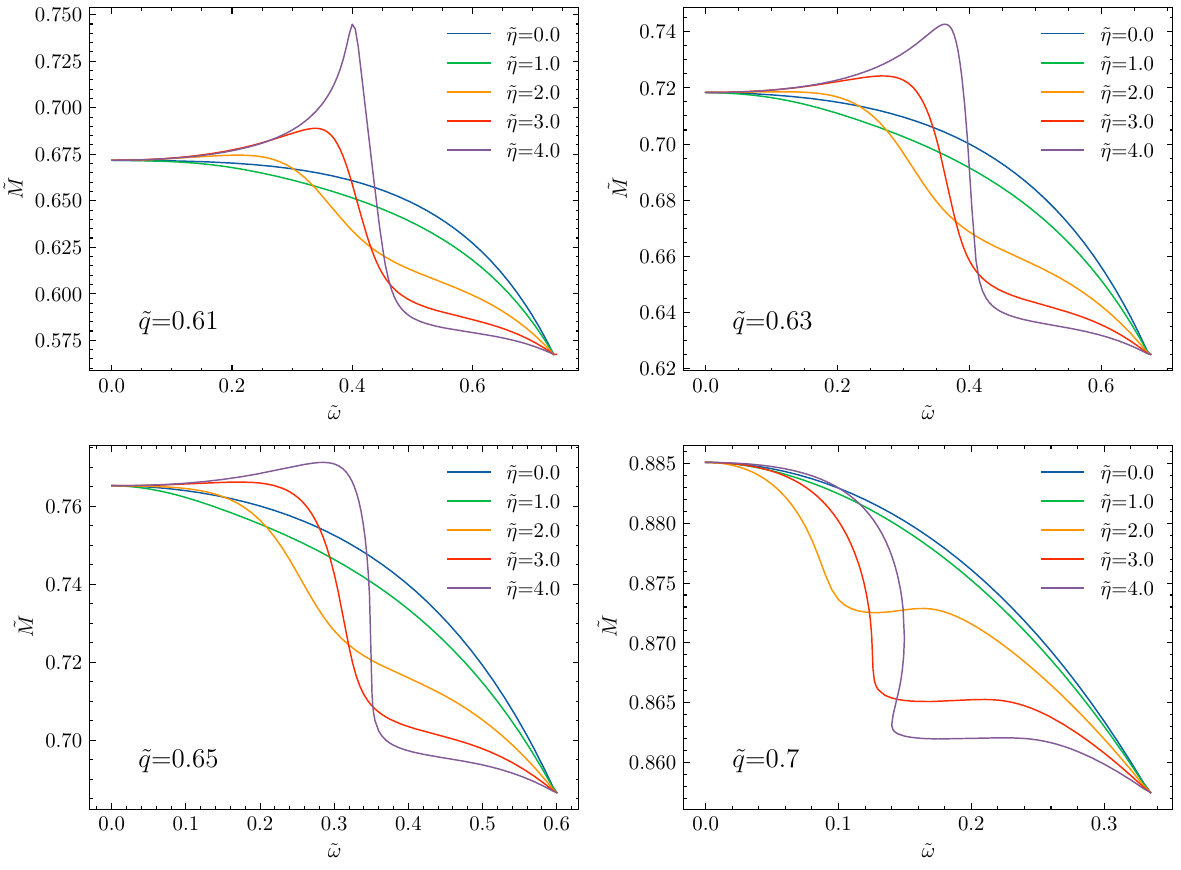}
\caption{$\tilde M$ versus $\tilde \omega$  with  different values of $\tilde \eta$ for $\tilde q > \tilde q_c$.}\label{figM2}
\end{figure}

\begin{figure}
\centering
\includegraphics[width=\linewidth]{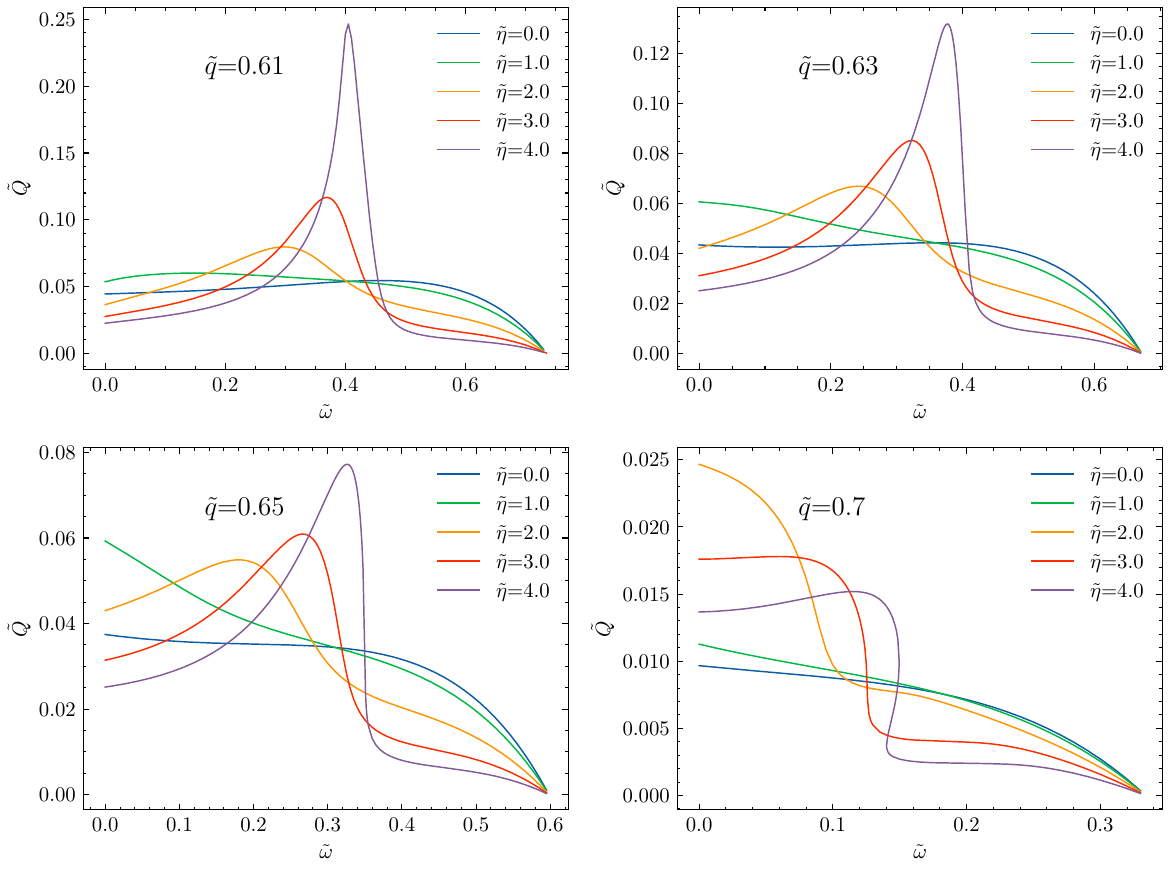}
\caption{ $\tilde Q$ versus $\tilde \omega$ with  different values of $\tilde \eta$ for $\tilde q > \tilde q_c$.}\label{figQ2}
\end{figure}

In this case, the shapes of the $\tilde M$ versus $\tilde \omega$ and $\tilde Q$ versus $\tilde \omega$ curves differ significantly from those in the $\tilde q < \tilde q_c$ case, with the most notable difference being the absence of the inflection point at the left end of the curves. For example, in Fig.~\ref{figM2} and Fig.~\ref{figQ2}, we present diagrams of the $\tilde M$ versus $\tilde \omega$ and $\tilde Q$ versus $\tilde \omega$ curves, respectively, for $\tilde q=0.61,0.63, 0.65,  0.7$. It is evident that for all values of $\tilde \eta$, there are no inflection points at the left end of the curves. Furthermore, the $\tilde M$ versus $\tilde \omega$ curves converge to the same point at both the left and right endpoints, with the values of $\tilde \omega$ corresponding to the left endpoints being $0.0001$ for all curves.

\begin{figure}

\subfigure[]{
\centering
\includegraphics[width=0.47\linewidth]{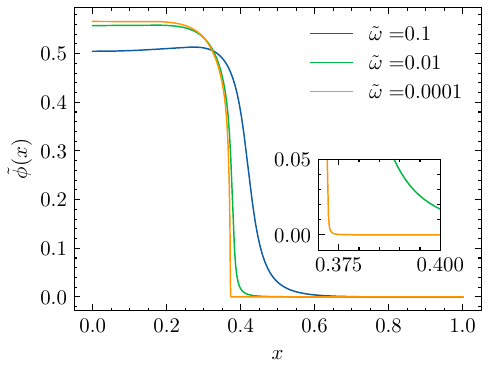}\label{q0.63-0.7_eta1.0-1.0_w0.0001-0.0001_-phi_part}
} 
\subfigure[]{
\centering
\includegraphics[width=0.47\linewidth]{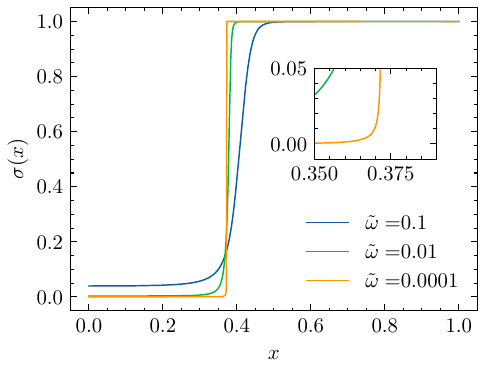}\label{q0.63-0.7_eta1.0-1.0_w0.0001-0.0001_-sigma_part}
}

\subfigure[]{
\includegraphics[width=0.47\linewidth]{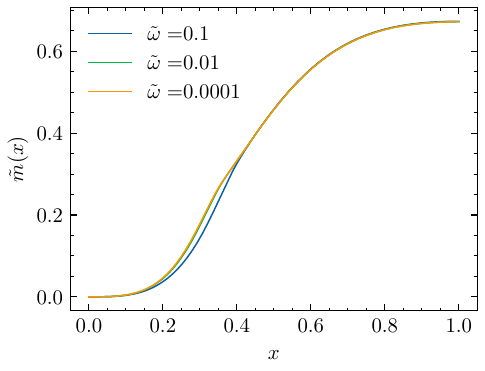}\label{q0.63-0.7_eta1.0-1.0_w0.0001-0.0001_-m_part}
}

\caption{Solutions of $\tilde \phi$ and $\sigma$ with $\tilde q = 0.61$ and $\tilde\eta = 3$.}\label{frozen}
\end{figure}

As previously mentioned, it is theoretically possible to obtain solutions with smaller values of $\tilde \omega$ { by refining the mesh  and decreasing the step size of $\tilde \omega$}. However, the stiffness of the problem increases as $\tilde \omega$ decreases. As shown in Fig.~\ref{frozen}, with $\tilde q =0.61$, $\tilde \eta =3$, and $\tilde \omega=0.1,0.01,0.0001$, the slopes of the solutions for the functions $\tilde \phi$ and $\sigma$ increase dramatically with decreasing $\tilde \omega$ around the point $x=0.35$. This indicates that the smaller the value of $\tilde \omega$, the more difficult the equations are to solve. Therefore, in this paper, we limit the minimum value of $\tilde \omega$ to $0.0001$.

\begin{figure}
\centering
\includegraphics[width=0.47\linewidth]{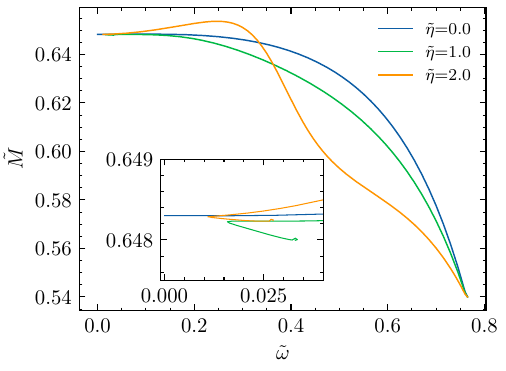}
\caption{$\tilde M$ versus $\tilde \omega$ with $\tilde q= 0.6$. The minimum frequencies corresponding to the curves with $\tilde \eta = 0$, $\tilde \eta = 1.0$, and $\tilde \eta = 2.0$ are $\tilde \omega = 0.0001$, $\tilde \omega = 0.01586$, and $\tilde \omega = 0.01104$, respectively.}\label{fig7}
\end{figure}

In Table~\ref{tab2}, there is a very slight difference in $\tilde q_c$ for different values of $\tilde \eta$, but it cannot be ignored. For example, the value of $\tilde q = 0.6$ is the critical value for $\tilde \eta = 0$, but not for $\tilde \eta = 1$ and $\tilde \eta = 2$. For $\tilde \eta = 0$, the lowest frequency value is $0.0001$, whereas for $\tilde \eta = 1$ and $\tilde \eta = 2$, the lowest frequency values are $0.01586$ and $0.01104$, respectively. Consequently, as shown in Fig.~\ref{fig7}, the $\tilde M$ versus $\tilde \omega$ curves corresponding to $\tilde \eta = 1$ and $2$ exhibit an inflection point at the left end.

\begin{figure}
\subfigure[]{
\centering
\includegraphics[width=0.47\linewidth]{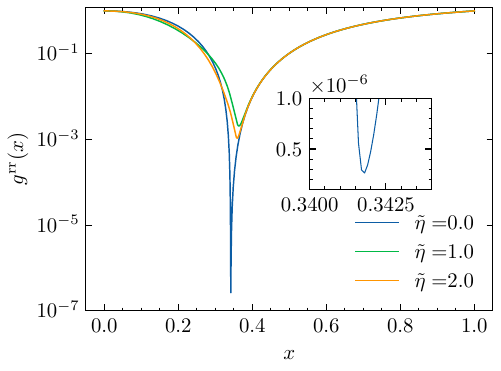}\label{fig81}
} 
\subfigure[]{
\centering
\includegraphics[width=0.47\linewidth]{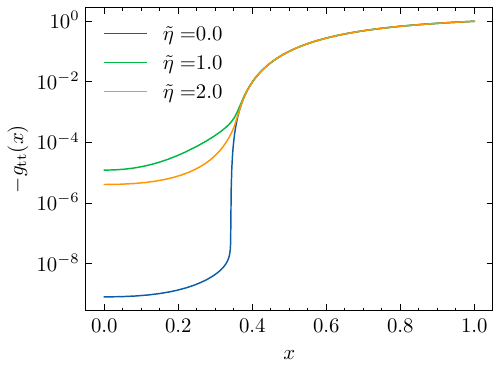}\label{fig82}
}
\caption{ Solutions of $g^{\mathrm{rr}}(x)$ and $-g_{\mathrm{tt}}(x)$ with $\tilde q= 0.6$. The frequencies corresponding to the curves with $ \tilde{\eta} = 0 $, $ \tilde{\eta} = 1.0 $, and $ \tilde{\eta} = 2.0 $ are $ \tilde{\omega} = 0.0001 $, $ \tilde{\omega} = 0.01586 $, and $ \tilde{\omega} = 0.01104 $, respectively.}\label{toforzen}
\end{figure}

Another consequence is the effect on the metrics. To illustrate this effect, we present the images of the metric components $g^{rr}$ and $-g_{tt}$ in Fig.~\ref{toforzen}. From these images, it is evident that the minima of $g^{rr}$ and $-g_{tt}$ at $\tilde \eta = 0$ are approximately three orders of magnitude smaller  than the minima of $g^{rr}$ and $-g_{tt}$ at $\tilde \eta = 1$ and $2$. { Moreover, compared to the cases $\tilde \eta = 1$ and $\tilde \eta = 2$, the minima of $g^{rr}$ and $-g_{tt}$ at $\tilde \eta = 0$ can be further reduced by refining the mesh  and decreasing the step size of $\tilde \omega$.}

\subsection{Frozen states and critical horizon} \label{FZS}
 
\begin{figure} 
\subfigure[]{
\centering
\includegraphics[width=0.47\linewidth]{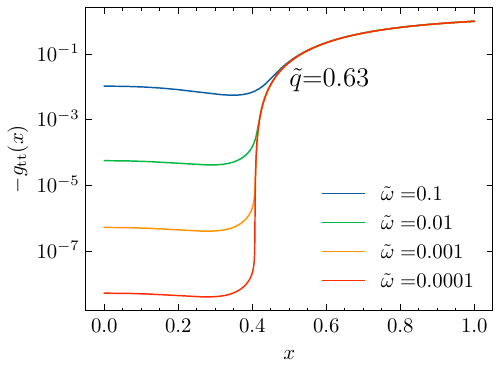} \label{fig61}
}
\subfigure[]{
\centering
\includegraphics[width=0.47\linewidth]{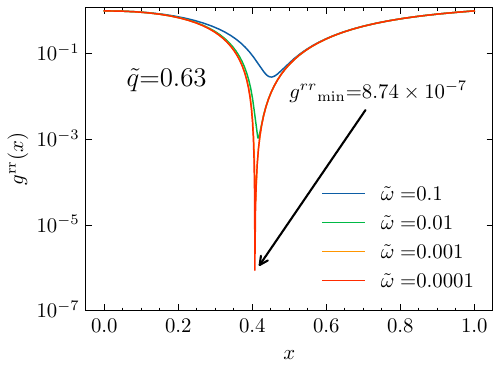}\label{fig62}
}

\subfigure[]{
\centering
\includegraphics[width=0.47\linewidth]{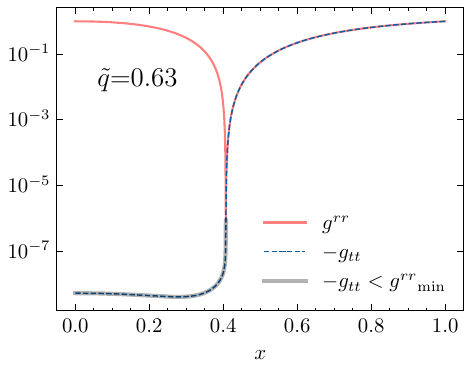}\label{fig63}
}
\subfigure[]{
\centering
\includegraphics[width=0.47\linewidth]{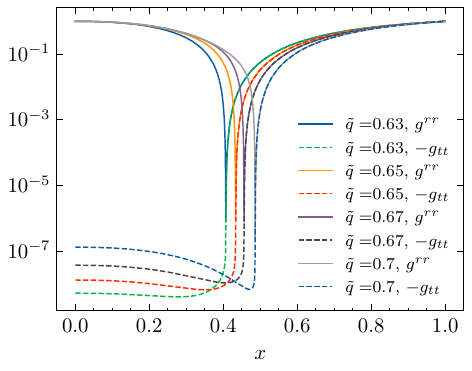}\label{fig64}
}

\caption{Solutions of $-g_{tt}$ and $g^{rr}$ ,  the vertical axes of are represented on a logarithmic scale, and   $\tilde \eta = 1$.   }\label{x_c}

\end{figure}

 In the case of $\tilde q > \tilde q_c$, as $\tilde \omega \to 0$, the value of $-g_{tt}$ also tends to zero in the central region inside the BBSs. As illustrated in Fig.~\ref{fig61}, within this region, the value of $-g_{tt}$ can be reduced by an order of magnitude, reaching $10^{-7}$. The boundary of this region, denoted by $x_c$, is determined by the minimum value of $g^{rr}$, referred to as $g^{rr}{}_\mathrm{min}$. Notably, $g^{rr}{}_\mathrm{min}$ is also reduced by an order of magnitude, reaching $10^{-7}$, as shown in Fig.~\ref{fig62}.

Fig.~\ref{fig63} demonstrates that within the specified range where $x < x_c$, $-g_{tt}$ is less than $g^{rr}{}_\mathrm{min}$. This phenomenon is universal, as evidenced by Fig.~\ref{fig64}. Consequently, when the minimum value of $g^{rr}$ reaches the order of $10^{-7}$, the sphere defined by the radial coordinate $x = x_c$ is termed the ``{\it critical horizon}'' \cite{YueWang2023a, HuangSunWang2023, ChenWang2024}.

This implies that, from the perspective of an observer at infinity, time nearly halts inside the critical horizon, and matter approaching the surface of the star appears to stop near the critical horizon. Thus, the BBSs are identified as frozen at this moment.


To gain an intuitive understanding of the critical horizon radius, we compare it with the event horizon radius of the magnetic Reissner-Nordstr\"om (RN) black hole. The line element of the magnetic RN black hole is given by

\begin{equation}\label{g_RN}
ds^2 = -\Delta(\tilde{r}) dt^2 + \Delta(\tilde{r})^{-1} d\tilde{r}^2 + \tilde{r}^2 \left(d\theta^2 + \sin^2\theta d\varphi^2\right),
\end{equation}
where
\[
\Delta(\tilde{r}) = 1 - \frac{2\tilde{M}}{\tilde{r}} + \frac{\tilde{q}^2}{\tilde{r}^2}.
\]
The event horizon radii of the magnetic RN black hole can be determined from the condition $\Delta = 0$:
\[
\tilde{r}^{\mathrm{H,RN}}_{\mathrm{inner/outer}} = \tilde{M} \mp \sqrt{\tilde{M}^2 - \tilde{q}^2} = \tilde{M} \left(1 \mp \sqrt{1 - \frac{\tilde{q}^2}{\tilde{M}^2}}\right),
\]
where $\tilde{r}^{\mathrm{H,RN}}_{\mathrm{inner}}$ and $\tilde{r}^{\mathrm{H,RN}}_{\mathrm{outer}}$ denote the inner and outer event horizons of the magnetic Reissner-Nordstr\"om (RN) black hole, respectively.

Our calculation results are summarized in Table~\ref{tab1}, where the critical horizon radius of a BBS is denoted by $\tilde{r}^{\mathrm{H}}_{c}$. From the table, we identify two key trends: (1) the ADM mass of the frozen BBSs increases monotonically with $\tilde{q}$; and (2) $\tilde{r}_{c}$ consistently satisfies the inequality $\tilde{r}^{\mathrm{H,RN}}_{\mathrm{inner}} < \tilde{r}^{\mathrm{H}}_{c} < \tilde{r}^{\mathrm{H,RN}}_{\mathrm{outer}}$. Additionally, similar to $\tilde{r}^{\mathrm{H,RN}}_{\mathrm{outer}}$, $\tilde{r}^{\mathrm{H}}_{c}$ exhibits a monotonic increase with $\tilde{q}$.

\begin{table}[tbp]
\centering
\begin{tabular}{lrccc}
\toprule
   & $\tilde M$  & $\tilde r^\mathrm{H}_c$ &$\tilde r^\mathrm{H,RN}_{\mathrm{outer}}$ &  $\tilde r^\mathrm{H,RN}_{\mathrm{inner}}$  \\
\midrule
$\tilde q = 0.61$ &0.67 & 0.59 & 0.95   & 0.39  \\
$\tilde q = 0.63$ &0.72 & 0.69 & 1.06   & 0.37 \\
$\tilde q = 0.65$ &0.77 & 0.77 & 1.17   & 0.36  \\
$\tilde q = 0.67$ &0.81 & 0.84 & 1.27   & 0.35   \\
$\tilde q = 0.70$ &0.86 & 0.98 & 1.43   & 0.34    \\
\bottomrule
\end{tabular}
\caption{ Radii of the critical horizons of  frozen BBSs and the event horizons of  magnetic RN block holes. All numerical values are rounded to two decimal places.}\label{tab1}
\end{table}

 \subsection{Compactness} \label{Com}

Each $\tilde{M}$ versus $\tilde{\omega}$ curve exhibits a maximum value of $\tilde{M}$, denoted by $\tilde{M}_{\mathrm{max}}$. The compactness discussed in this section is specifically evaluated at this maximum mass value.

The definition of compactness adopted here is presented in Ref.~\cite{CanoMachetMyin2024}. The radius $X$ is determined by solving the equation:
\begin{equation}
\frac{\tilde{m}(X)}{\tilde{M}_{\mathrm{max}}} = 0.99.
\end{equation}
Additional definitions of compactness can be found in Ref.~\cite{CollodelDoneva2022}. Using Eq.~\eqref{x_r}, the corresponding physical radius $\tilde{R}$ can be derived as
\begin{equation}\label{r_x}
\tilde{R} = \frac{X}{1 - X},
\end{equation}
and the compactness is then calculated by
\begin{equation}
\mathcal{C} = \frac{\tilde{M}_{\mathrm{max}}}{\tilde{R}}.
\end{equation}
Simultaneously, the frequency corresponding to $\tilde{M}_{\mathrm{max}}$, denoted by $\tilde{\omega}^*$, can also be identified.

\begin{figure}[htbp]

\subfigure[]{
\centering
\includegraphics[width=0.47\linewidth]{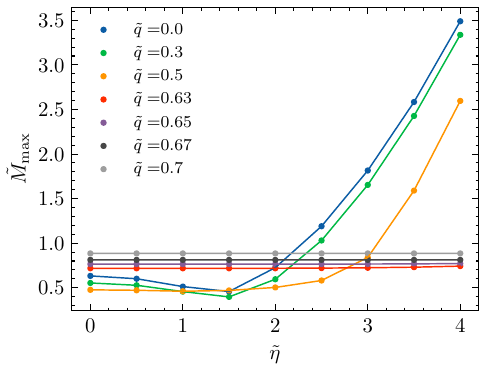}  \label{fig9a}
}
\subfigure[]{
\centering
\includegraphics[width=0.47\linewidth]{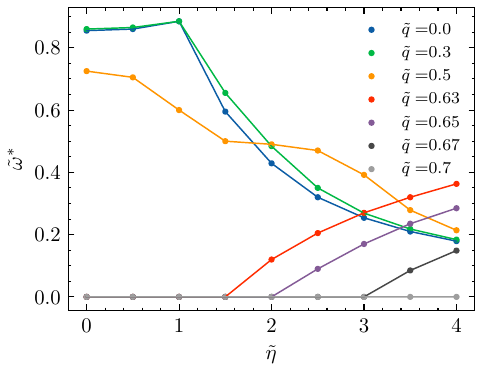} \label{fig9b}
}
\caption{(a) $\tilde{M}_{\mathrm{max}}$ versus $\tilde{\eta}$ with $\tilde{s} = 0.2$. (b) $\tilde{\omega}^*$ versus $\tilde{\eta}$ with $\tilde{s} = 0.2$. The dots in the figure correspond to the calculated values of $\tilde{M}_{\mathrm{max}}$ and $\tilde{\omega}^*$.} \label{fig9}
\end{figure}

Fig.~\ref{fig9} illustrates the variations of  $\tilde{M}_{\mathrm{max}}$ and  $\tilde{\omega}^*$ as functions of $\tilde{\eta}$. For the case of $\tilde{q} < \tilde{q}_c$, as $\tilde{\eta}$ increases, $\tilde{M}_{\mathrm{max}}$ initially decreases and then increases, with the maximum difference in $\tilde{M}_{\mathrm{max}}$ exceeding 2.0. When $\tilde{\eta} > 1$, the value of $\tilde{\omega}^*$ decreases monotonically with increasing $\tilde{\eta}$. On the other hand, for $\tilde{q} > \tilde{q}_c$, $\tilde{M}_{\mathrm{max}}$ remains nearly constant as $\tilde{\eta}$ increases, and the maximum difference in $\tilde{M}_{\mathrm{max}}$ is less than 0.1. Additionally, except for the case of $\tilde{q} = 0.7$ where $\tilde{\omega}^*$ remains constant, $\tilde{\omega}^*$ increases monotonically with $\tilde{\eta}$ beyond a certain value of $\tilde{\eta}$.

\begin{figure}[htbp]
\subfigure[]{
\centering
\includegraphics[width=0.47\linewidth]{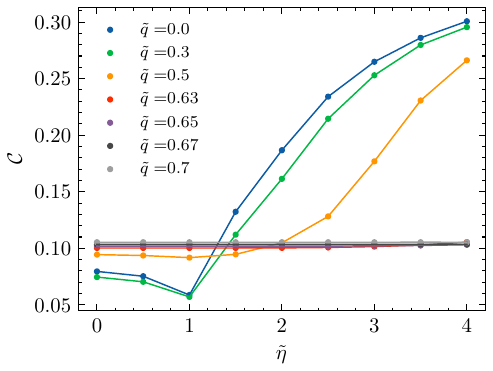}  \label{fig10a}
}
\subfigure[]{
\centering
\includegraphics[width=0.47\linewidth]{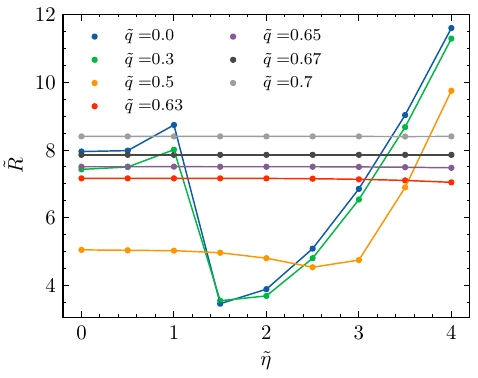} \label{fig10b}
}

\caption{  (a) Compactness $\mathcal{C}$ versus $\tilde{\eta}$. (b) Radius $\tilde{R}$ versus $\tilde{\eta}$. The dots in the figure correspond to the calculated values of $\mathcal{C}$ and $\tilde{R}$.} \label{fig10}
\end{figure}

Fig.~\ref{fig10} depicts the compactness $\mathcal{C}$ and the radius $\tilde{R}$ of BBSs as functions of $\tilde{\eta}$. For $\tilde{q} < \tilde{q}_c$, the compactness $\mathcal{C}$ decreases initially and subsequently increases as $\tilde{\eta}$ increases, whereas the radius $\tilde{R}$ initially exhibits fluctuations and then undergoes a significant increase overall. In contrast, for $\tilde{q} > \tilde{q}_c$, both $\mathcal{C}$ and $\tilde{R}$ demonstrate negligible variation with increasing $\tilde{\eta}$, with the compactness values remaining consistently around $\mathcal{C} \approx 0.1$.

\subsection{Light Rings} \label{LRS}

The geodesic equation for photons in a gravitational field is given by
\begin{equation}\label{geodesic}
g_{\mu\nu} \dot{x}^{\mu} \dot{x}^{\nu} = 0,
\end{equation}
where the dot denotes a derivative with respect to an affine parameter along the geodesic and $x^{\mu} = (t, r, \theta,\phi)$.

Due to the static spherical symmetry, we assume photon orbits lie in the equatorial plane $\theta = \pi/2$, implying $\dot{\theta} = 0$. The metric \eqref{equ10} possesses two Killing vectors $\partial_t$ and $\partial_\varphi$, corresponding to conserved quantities: the energy $E = - g_{tt} \dot{t}$ and angular momentum $L = \tilde r^2 \dot{\varphi}$. Equation~\eqref{geodesic} then reduces to 
\begin{equation}\label{geodesic1}
\frac{1}{2} \dot{\tilde r}^2 + V(\tilde r) = 0, 
\end{equation}
with the effective potential \cite{deSaLimaHerdeiro2024}
\begin{equation}\label{potential1}
V(\tilde r) = \frac{1}{2} \frac{L^2}{-g_{tt}g_{11}} \left( \frac{-g_{tt}}{\tilde r^2} - \frac{1}{b^2} \right),
\end{equation}
where $b = L/E$ is the impact parameter. This formulation describes a classical particle of unit mass moving in the potential $V(\tilde r)$ with zero energy \cite{Carroll2014a}.

Light rings (LRs) correspond to circular photon orbits satisfying $V(\tilde r) = 0$ and $V'(\tilde r) = 0$, yielding:
\begin{align}
\left( \frac{-g_{tt}}{\tilde r^2} - \frac{1}{b^2} \right) &= 0, \label{lr_cond1} \\
\left( -\frac{g_{tt}}{\tilde r^2} \right)' &= 0. \label{lr_cond2}
\end{align}
One can define the reduced effective potential:
\begin{equation}\label{potential3}
V_{\mathrm{eff}}(\tilde r) = -\frac{g_{tt}}{\tilde r^2}
\end{equation}
to characterize LR positions. Similarly, the radial stability condition $V''(r)>0$  can be replaced by $V_{\mathrm{eff}}''(\tilde r) >0$. The stability of the orbit along the $\theta$-direction has been rigorously proved in Ref.~\cite{CunhaBertiHerdeiro2017}.

\begin{figure}[htbp]
\subfigure[]{
\centering
\includegraphics[width=0.47\linewidth]{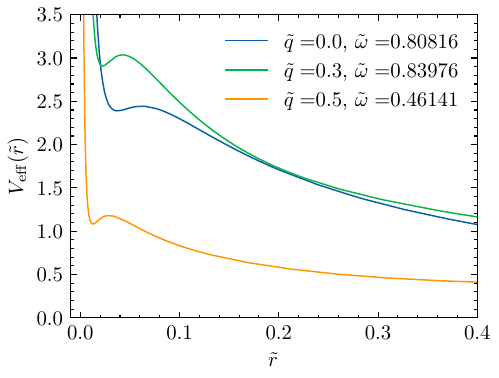} 
\label{LR-1a}
}
\subfigure[]{
\centering
\includegraphics[width=0.47\linewidth]{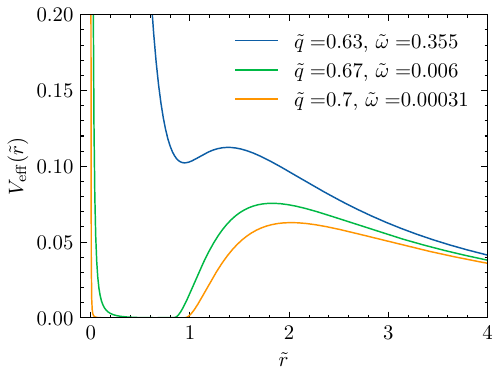} 
\label{LR-1b}
}
\caption{Effective potentials for photon circular orbits with $\tilde\eta = 1$ and varying $\tilde{q}$, $\tilde{\omega}$.} 
\label{LR-1}
\end{figure}

Our calculation reveals  LR solutions exist for all parameter combinations:
\begin{equation*}
\tilde q \in \{0.0, 0.1, 0.3, 0.5, 0.6, 0.61, 0.63, 0.65, 0.67, 0.7 \}, 
\end{equation*}
and
\begin{equation*}
 \tilde\eta \in \{0.0,0.5,1.0,1.5,2.0,3.0,3.5,4.0 \}.
\end{equation*}
Fig.~\ref{LR-1} displays representative effective potentials for $\tilde\eta = 1$ with varying $\tilde q$ and $\tilde\omega$.

\begin{figure}[htbp]
\subfigure[]{
\centering
\includegraphics[width=0.47\linewidth]{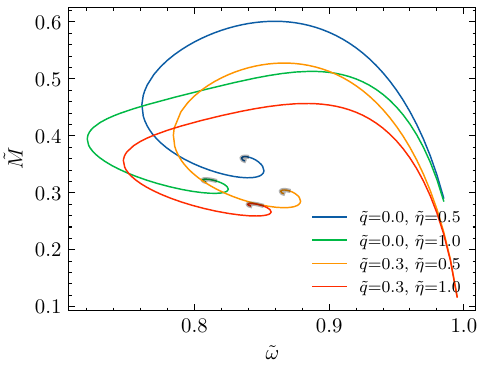}  
\label{LR-21}
}
\subfigure[]{
\centering
\includegraphics[width=0.47\linewidth]{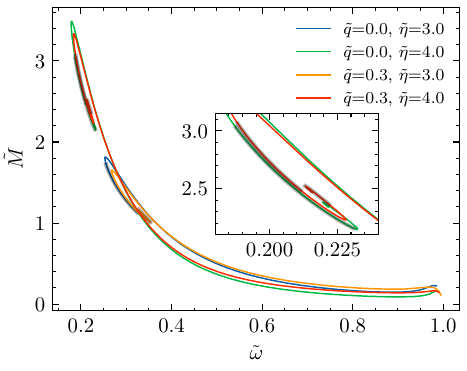}  
\label{LR-22}
}
\subfigure[]{
\centering
\includegraphics[width=0.47\linewidth]{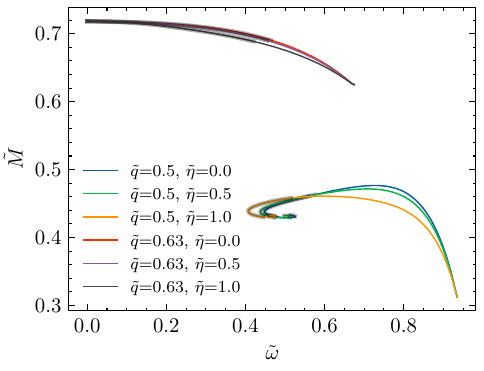}  
\label{LR-23}
}
\subfigure[]{
\centering
\includegraphics[width=0.47\linewidth]{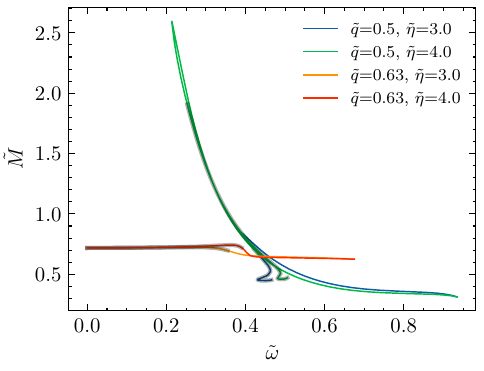}  
\label{LR-24}
}
\caption{$\tilde M$-$\tilde\omega$ relations with LR regions (gray) for various $\tilde q$, $\tilde\eta$.} 
\label{LR-2}
\end{figure}

LR solutions exist only within specific frequency ranges. Fig.~\ref{LR-2} illustrates this by shading LR regions on $\tilde M$-$\tilde\omega$ curves. For small $\tilde q$ and $\tilde\eta$, LRs primarily occupy counterclockwise-end regions (subfigure~\ref{LR-21}). As these parameters increase, LR regions expand significantly (subfigures~\ref{LR-22}-\ref{LR-24}).

Comparing frozen BBSs with magnetic RN black holes, the RN effective potential is:
\begin{equation}\label{pRN}
{V}_{\mathrm{eff}}^{\mathrm{RN}}(\tilde r) = -\frac{\Delta(\tilde r)}{\tilde r^2},
\end{equation}
yielding two LRs at:
\begin{align}
\tilde r_\mathrm{inner}^{\mathrm{LR,RN}} &= \frac{1}{2} \left(3 \tilde{M} - \sqrt{9 \tilde{M}^2 - 8 \tilde q^2}\right), \\
\tilde r_\mathrm{outer}^{\mathrm{LR,RN}} &= \frac{1}{2} \left(3 \tilde{M} + \sqrt{9 \tilde{M}^2 - 8 \tilde q^2}\right).
\end{align}

\begin{figure}[htbp]
\centering
\includegraphics[width=\linewidth]{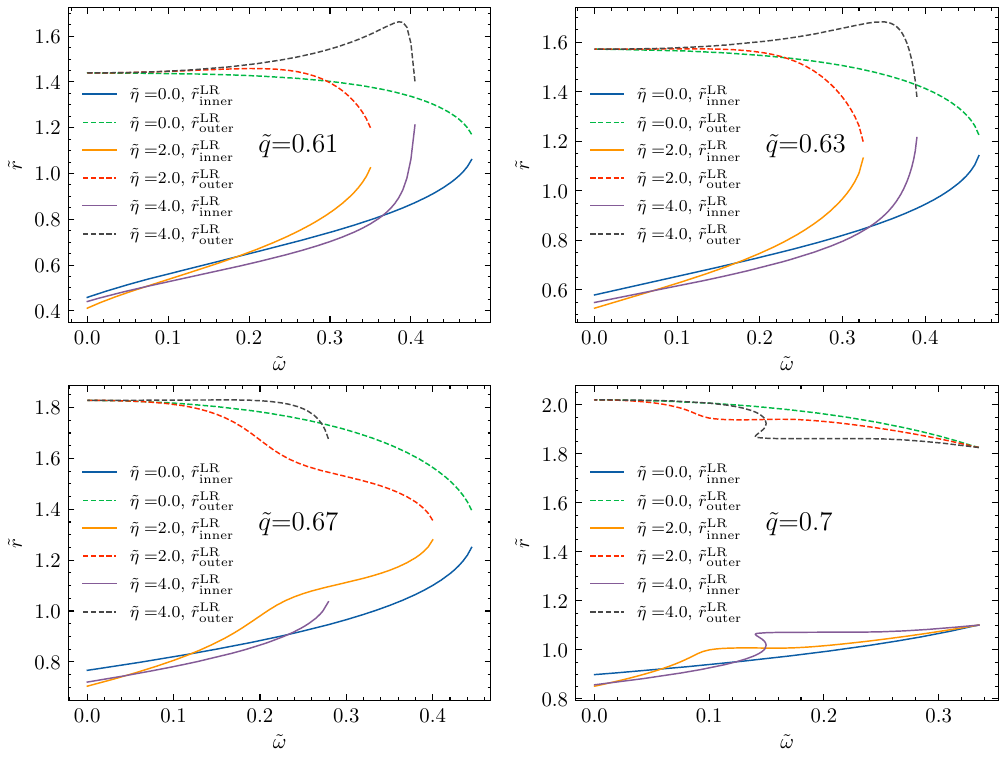}  
\caption{Inner ($\tilde r_\mathrm{inner}^{\mathrm{LR}}$) and outer ($\tilde r_\mathrm{outer}^{\mathrm{LR}}$) LR radii versus $\tilde\omega$ for BBSs.} 
\label{LRs}
\end{figure}

Boson stars also exhibit two light rings (LRs) \cite{MacedoPaniCardoso2013}, with the inner and outer radii denoted by $\tilde{r}_\mathrm{inner}^\mathrm{LR}$ and $\tilde{r}_\mathrm{outer}^\mathrm{LR}$, respectively. Fig.~\ref{LRs} plots the radii of the LRs of BBSs as functions of $\tilde{\omega}$ for  $\tilde{q} = 0.61, 0.63, 0.67, 0.7$. 
As illustrated in Fig.~\ref{LRs}, as $\tilde{\omega} \to 0$, the outer radius $\tilde{r}_\mathrm{outer}^\mathrm{LR}$ converges to a fixed value, while the inner radius $\tilde{r}_\mathrm{inner}^\mathrm{LR}$ does not. This convergence of the outer radius to a fixed point implies that the position of the outer LRs of frozen BBSs is independent of $\tilde{\eta}$.

\begin{figure}[htbp]
\centering
\includegraphics[width=\linewidth]{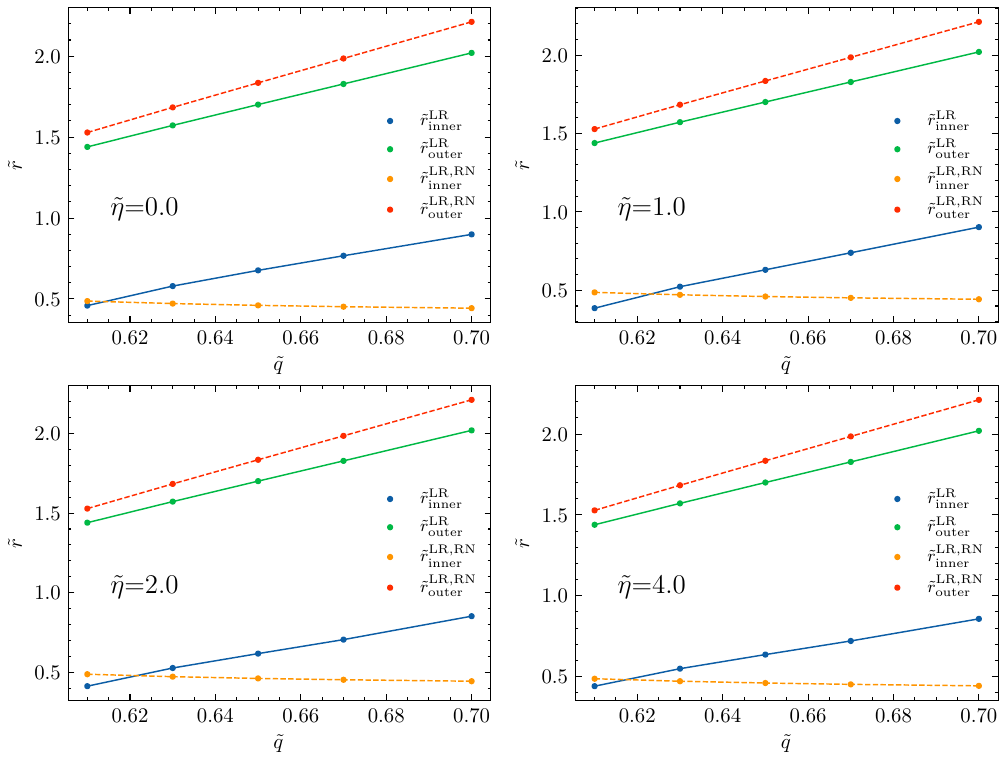} 
\caption{Comparison of LR radii between frozen BBSs and RN black holes versus $\tilde q$.} 
\label{LR-4}
\end{figure}

Fig.~\ref{LR-4} illustrates the relationship between the light ring radii of frozen BBSs and RNs black holes for  $\tilde{\eta} = 0.0, 1.0, 2.0, 4.0$. For the tested values of $\tilde{q}$, the outer light ring radius of BBSs, $\tilde{r}_\mathrm{outer}^\mathrm{LR}$, is consistently smaller than that of RN black holes, $\tilde{r}_\mathrm{outer}^\mathrm{LR,RN}$. Conversely, the inner light ring radius of BBSs, $\tilde{r}_\mathrm{inner}^\mathrm{LR}$, is larger than that of RN black holes, $\tilde{r}_\mathrm{inner}^\mathrm{LR,RN}$, for all tested values of $\tilde{q}$ except $\tilde{q} = 0.61$. Additionally, both $\tilde{r}_\mathrm{inner}^\mathrm{LR}$ and $\tilde{r}_\mathrm{outer}^\mathrm{LR}$ increase with increasing $\tilde{q}$.

\section{Conclusions}\label{Conc}

In this work, we have investigated a Bardeen model that couples Einstein's gravity with nonlinear electromagnetic fields and non-topological soliton complex scalar fields. When the magnetic charge $\tilde{q} = 0$ and the complex scalar field  self-interaction parameter $\tilde{\eta} = 0$, the model reduces to the mini-boson stars model, serving as a baseline for our study. To explore the effects of magnetic charge and scalar field self-interaction, we systematically increase the values of $\tilde{q}$ and $\tilde{\eta}$. We pay particular attention to parameter values leading to frozen star solutions.

Our findings reveal that the magnetic charge $\tilde{q}$ plays a crucial role in determining the frozen states of Bardeen boson stars (BBSs), with different critical values for varying $\tilde{\eta}$. Notably, the ADM mass $\tilde{M}$ of frozen BBSs is independent of $\tilde{\eta}$, and these frozen solutions exhibit a critical horizon whose radius lies between the inner and outer horizon radii of a magnetic Reissner-Nordström (RN) black hole.

For fixed $\tilde q$ and $\tilde{\eta}$,  we have determined the maximum ADM mass of BBSs from the $\tilde{M}$-$\tilde{\omega}$ curves and computed the corresponding compactness. Notably, for $\tilde{q} > \tilde{q}_c$, both the compactness and the maximum $\tilde{M}$ remain nearly constant with respect to $\tilde\eta$.

We also investigate the existence of light rings for BBSs. Our calculations demonstrate  that light ring solutions exist for all tested combinations of $\tilde{q}$ and $\tilde{\eta}$, and all frozen stars possess light rings. The outer light ring radius of frozen stars is independent of $\tilde{\eta}$. Compared to magnetic RN black holes, frozen BBSs exhibit a smaller outer light ring radius.

Furthermore, we have examined the existence of light rings (LRs) for BBSs. Our results demonstrate that LR solutions exist for all tested combinations of $\tilde{q}$ and $\tilde{\eta}$, and all frozen BBSs possess LRs. The outer LR radius of frozen stars is independent of $\tilde{\eta}$, and compared to magnetic RN black holes, frozen BBSs exhibit a smaller outer LR radius. 

%

\section{Acknowledgment}
This work is supported by the National Natural Science Foundation of China (Grant No. 12275110  and No. 12247101)  and  the National Key Research and Development Program of China (Grant No. 2022YFC2204101 and 2020YFC2201503).

\bibliographystyle{JHEP}

\providecommand{\href}[2]{#2}\begingroup\raggedright\endgroup

\end{document}